\documentclass[12pt,preprint]{aastex}

\shorttitle{Pfleiderer\,2: a new globular cluster in the Galaxy}
\shortauthors{Ortolani et al.}

\begin{document}


\title{Pfleiderer\,2: identification of a new globular cluster in the Galaxy}

\author{S. Ortolani\altaffilmark{1}}
\affil{Universit\`a di Padova, Dipartimento di Astronomia, Vicolo
 dell'Osservatorio 5, I-35122 Padova, Italy}
\email{ortolani@pd.astro.it}
\and
\author{C. Bonatto, E. Bica}
\affil{Universidade Federal do Rio Grande do Sul, 
Departamento de Astronomia, CP 15051, Porto Alegre 91501-970, Brazil}
\email{charles@if.ufrgs.br, bica@if.ufrgs.br}
\and
\author{B. Barbuy}
\affil{Universidade de S\~ao Paulo, Departamento de Astronomia, 
 Rua do Mat\~ao 1226, S\~ao Paulo 05508-900, Brazil}
\email{barbuy@astro.iag.usp.br}


\altaffiltext{1}{Based on observations made with the italian Telescopio
Nazionale Galileo (TNG) operated on the island of La Palma, by the
Fundaci\'on Galileo Galilei of INAF (Istituto Nazionale di Astrofisica)
at the spanish Observatorio del Roque de los Muchachos of the
Instituto de Astrofisica de Canarias}

\begin{abstract}
{\bf We provide evidence that indicate the star cluster Pfleiderer\,2, 
which is projected in a rich field, as a newly identified Galactic globular 
cluster.} Since it is located in a crowded field, core extraction and 
decontamination tools were applied to reveal the cluster sequences in B, V 
and I Color-Magnitude Diagrams (CMDs). The main CMD features of Pfleiderer\,2 
are a tilted Red Giant Branch, and a red Horizontal Branch, indicating a high 
metallicity around solar. The reddening is E(B-V)=1.01. The globular cluster 
is located at a distance from the Sun d$_{\odot}$ = 16$\pm$2 kpc.

The cluster is located at 2.7 kpc above the Galactic plane and at a distance 
from the Galactic center of  R$_{\rm GC}$=9.7\,kpc, which is unusual for a 
metal-rich globular cluster. 

\end{abstract}

\keywords{globular clusters: individual: Pfleiderer\,2 --   HR diagram}

\section{Introduction}

Classical globular clusters are prominent objects with respect to the surrounding 
fields. In recent years, however, low luminosity globular clusters have been
identified or discovered in the Galaxy. These include obscured or field contaminated 
objects, or a combination of these factors (e.g. \citealt{OBB06}, and references 
therein). \citet{Kurtev08} recently found GLIMPSE-C02, an extremely obscured and 
metal rich globular cluster, {\bf which was almost simultaneously detected by
\citet{StrKob08}.}

Pfleiderer\,2 was found by \citet{Pfl77} and has been also referred to as PWM2.
The cluster center  is located at (J2000) $\alpha = 17^{\rm h}58^{\rm m}40^{\rm s}$, 
$\delta = -5\degr04\arcmin30\arcsec$, with Galactic coordinates $\ell=22.28\degr$, 
$b=+9.32\degr$. The angular size is $\sim2.5\arcmin$, as seen on DSS/XDSS\footnote{Extracted 
from the Canadian Astronomy Data Centre (CADC) - \em http://cadcwww.dao.nrc.ca/} images.

In Sect.~\ref{Obs} the observations and reductions are described. In Sect.~\ref{Decont}
we describe the decontamination procedures. Colour Magnitude Diagrams (CMD) are presented 
and fundamental parameters derived in Sect.~\ref{CMD}. The cluster structure is studied 
in Sect.~\ref{Struc}. Concluding remarks are given in Sect.~\ref{Conclu}.
 
\section {Observations} 
\label{Obs}

\begin{figure}
\centerline{
\resizebox{10.0cm}{!}{\includegraphics[angle=0,scale=.50]{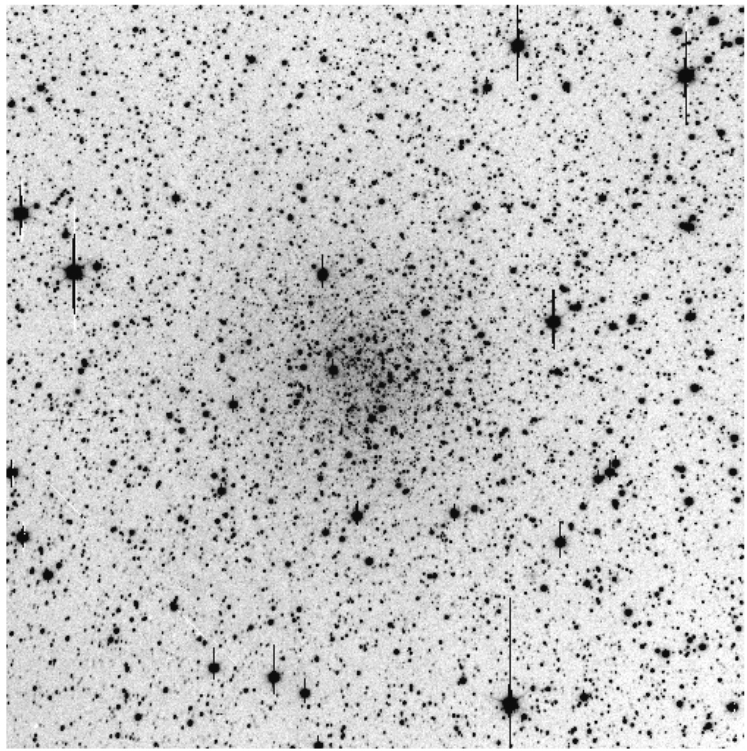}}}
\caption{Pfleiderer\,2 full field 420sec. I image. 
East is to the right and North to the top.}
\label{fig1}
\end{figure}

Pfleiderer\,2 was observed on 2008 June 5, 6,
with the 3.58m Galileo telescope (TNG)  at La Palma, equipped 
with the Dolores spectrograph focal reducer.
A EEV 4260 CCD detector  with $2048\times2048$ pixels, 
of pixel size 13 $\mu$m was used. A pixel corresponds 
to $0.252\arcsec$ on the sky, and the full 
field of the camera is $8.6\arcmin\times 8.6\arcmin$. 
The log of observations is provided in Table 1.
In Fig.~\ref{fig1} is shown a 420 sec full field  $I$ image of Pfleiderer\,2,
 showing a rather prominent cluster.

\begin{deluxetable}{rrrrrrrrr}
\tabletypesize{\scriptsize}
\tablecaption{Log of observations \label{log}}
\tablewidth{0pt}
\tablehead{
\colhead{Filter} & \colhead{Date} & \colhead{exp(sec)} & \colhead{seeing (\arcsec)} & 
\colhead{airmass} & 
}
\startdata           
V  &  5-6/06/2008  &     600    &       1.1    &    1.2      & \\
I  &      ``         &      30    &         1.2  &      1.2    & \\
I  &       ``        &     420    &        1.1   &     1.2     & \\
V  & 6-7/06/2008   &    30      &       1.6    &    1.7      & \\
V  &      ``         &   900      &     1.6      &  1.6        & \\
B  &       ``        &    60      &      1.5     &   1.5       & \\
B  &        ``       &    20      &      1.4     &   1.5       & \\
B  &         ``      &   900      &     1.4      &  1.4        & \\
B  &         ``      &    60      &     1.4      &  1.4        & \\
\enddata
\end{deluxetable}

\subsection{Pfleiderer\,2 calibrations}

The calibration was performed using 3 different Landolt (1983, 1992) standard fields 
observed in BVI, the first night  (PG1633+099, SA 113337, PG 2213-06) with different 
exposure times and in different positions. We checked shutter time dependence and zero 
point variations across the field. In the second night (June 6th) five additional 
\citet{Landolt92} fields have been observed (PG 1323-086, PG1528+062, PG1633+099, 
SA113 337, PG 2213-06). The airmass corrections of 0.15 in V, 0.25 in B and 0.07 in I,
have been done using the standard values given for La Palma in clear nights (with no 
dust in the atmosphere), available at  the DOLORES 
webpage\footnote{http://www.tng.iac.es/info/la$\_$palma$\_$sky.html}

The derived calibration equations are:

V = 28.44-0.06(V-I)+v

I = 28.28-0.07(V-I)+i

B = 28.59+0.12(B-V)+b

V = 28.43-0.06(B-V)+v

where the zero points are calculated for 10\,s exposures at 1.15 airmasses, with an aperture 
of $10\arcsec\times1\arcsec$. The bvi are instrumental magnitudes (counts in ADUs, gain 1 
e-/ADU). BVI are the calibrated magnitudes. The zero points were remarkably stable and identical 
in the two nights. No shutter time effects have been detected at the 0.01 mag. level in the exposure 
range of our images (from 1 to 30 s). In the $1000\times1000$ pixels central area the standard 
stars showed a very good uniformity, but their flux rapidly decreased at the frame corners. The 
repeated measurements of the standard stars across the field have been used to separate the 
contribution of the sky concentration from the flat field, including the decrease of sensitivity 
of the system (possibly due to some vignetting) at the edge. Measurements performed with different
apertures excluded the possibility that the loss of light at the edges was due to degradation of 
the PSF.

\subsection{Reductions}

\begin{figure}
\centerline{
\includegraphics[angle=0,scale=.50]{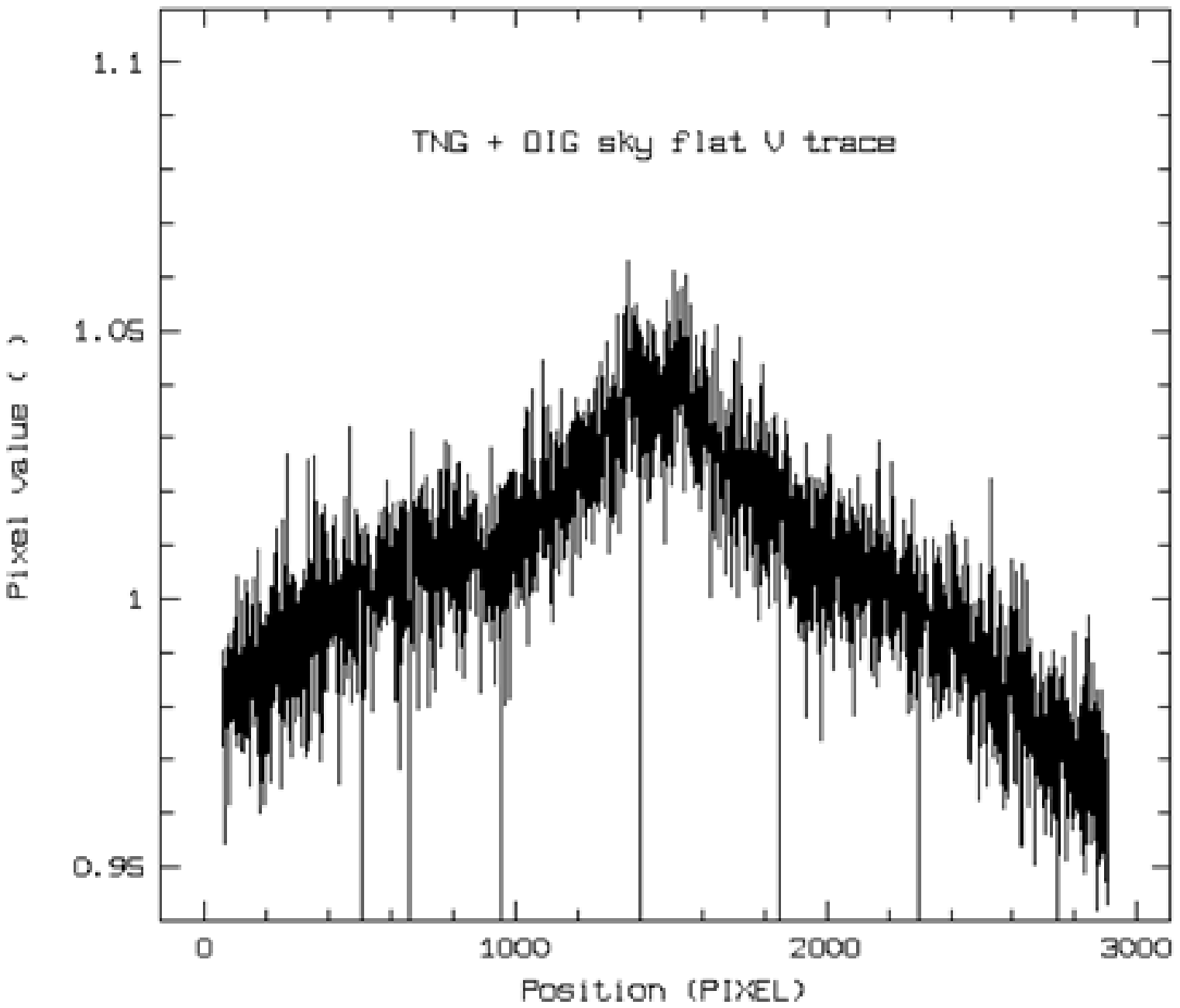}
\includegraphics[angle=0,scale=.50]{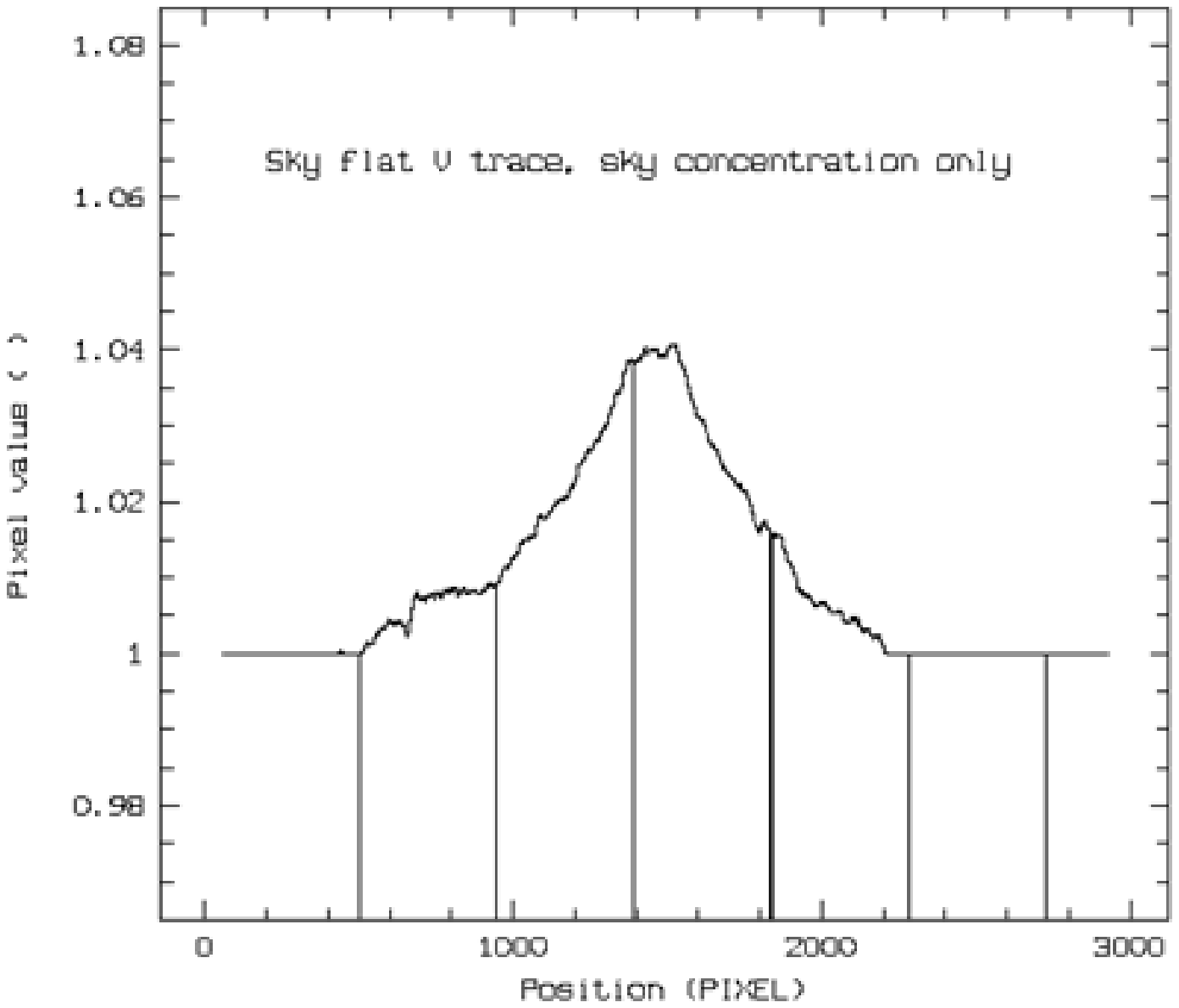}}
\caption{Left panel: V Flatfield (corner-to-corner) taken at twilight;
Right panel: corner-to-corner scan of the smoothed sky concentration,
removed from the original flatfield. }
\label{fig2}
\end{figure}

The CCD frame reductions were performed using a slightly more complex procedure than the standard 
flat field division. This was needed due to the analysis of the  standard star fluxes across the 
field, compared to the background brightness, and showed a sky flux excess in an area roughly 
corresponding to the central 500 pixels. This excess is interpreted as the effect called ``sky 
concentration'' (\citealt{AFS95}\footnote{ 
http://www.ls.eso.org/lasilla/Telescopes/2p2T.old.obsolete/D1p5M/RepsFinal/Final/gain$\_$calibration.ps}). 
This is known to be caused by internal reflections and it is color dependent. 
In our images it reached about 6\% in I, 5\% in V (Fig.~\ref{fig2}) and it was almost 
negligible in B. If the light concentration is not removed from the flatfields, it causes a 
systematic photometric error across the field of the same amount. The light loss at the edges 
was instead almost constant with the colors, up to 5\% from center to corner. The flatfielding 
(after bias correction) was performed in four main steps: (1) a map of sky concentration was 
prepared using the standard stars, compared to the sky level; (2) the contribution of the light 
concentration was extracted from the flatfields and smoothed (Fig.~\ref{fig2}); (3) the 
light concentration was scaled to the value of the sky images and subtracted; (4) the flatfield 
obtained after removing the light concentration was used to correct the high frequencies and 
sensitivity variations (including vignetting). 

The corrected images have been processed using 
Daophot II and Allstars (\citealt{Stetson87}; \citealt{Stetson94}). The software is available 
in MIDAS for the instrumental photometry of individual stars, and it was converted into the 
standard Johnson-Cousins system, defined from the Landolt stars, using the above calibration 
equations. The conversion from the instrumental (PSF convolved) magnitudes to the calibrated 
ones revealed an additional problem related to the image quality variation across the field. 
This produces a variable aperture correction, changing with time and filter. In Pfleiderer\,2 
images it was of the order of 0.02 magnitudes, with a peak around 0.04 magnitudes at the extreme 
corners. For this reason we adopted the aperture correction established from stars located in 
the central 200 pixels which is representative of most of the image within about 1\% to 2\%. 

The resulting calibration accuracy, taking into account the uncertainties in the calibration 
equations ($\sim$0.01 mag), the flatfield procedure (an additional 0.01 - 0.02 mag) and the 
aperture corrections (0.02 mag), is estimated to be within $\pm0.04$ in a single filter. 
The relative photometric errors have been derived from short exposures (15s in V and 10s in I) 
frame to frame images. The errors are 0.03 mag for 14 $<$ V $<$ 16.5, increasing to about 
0.2\,mag at V=20.5. In order to have the errors for our deep 600s exposures, they must be 
scaled to the corresponding magnitudes. The 600 sec images imply a factor 40 larger
in exposure times, and the corresponding error values can be estimated from the ones given
above.

\section{Decontamination procedures}
\label{Decont}

Field-stars are an important contaminant of CMDs in rich fields, especially near the disk and bulge.
The decontamination algorithm employed here is based on a three-dimensional routine designed for the
wide-field photometry of 2MASS data, as developed  in \citet{BB07}, \citet{ProbFSR} and \citet{F1603}.
In the present case we adapted the original algorithm to work with photometry obtained with a large 
telescope and a single color. For clarity, we recall the basic procedures. The algorithm divides the 
whole magnitude and color ranges into a grid of CMD cells. For a given cluster extraction and comparison 
field, it estimates the relative number densities of probable field and cluster stars present in each 
cell. The estimated number of field stars is subsequently subtracted from each cell. {\bf Decontamination 
results for different cluster/field contrasts are given and discussed in the above cited papers. The 
reference cell dimensions are $\rm\Delta\,mag=1.0$ and $\rm\Delta\,color=0.2$. However, to minimize 
spurious results, several runs of the decontamination procedure are used, with different input parameters. 

In the present paper, different cell sizes are considered, with $\rm\Delta\,mag$ and $\rm\Delta\,color$ 
taken from 0.5, 1.0 and 
2.0 times the reference values. Also, the cell grid is shifted by -1/3, 0 and +1/3 of the respective cell 
size in both the colour and magnitude axes. Taking together all the grid/cell size setups, we are left 
with 81 different and independent decontamination combinations. Stars are ranked according to the number 
of times they survive each run. Finally, only the highest ranked stars are considered as cluster members 
and transposed to the respective decontaminated CMD. In the case of Pfleiderer\,2, the decontaminated 
stars have a survival frequency above $S_d=70\%$. We note that the two bright and red stars (Figs.~\ref{fig6} 
and \ref{fig7}) are among those with $S_d=100\%$, which implies that the observed field is scarcely populated
of stars with similar colour and magnitude.}

\begin{figure}
\centerline{
\resizebox{13.0cm}{!}{\includegraphics[angle=0,scale=.50]{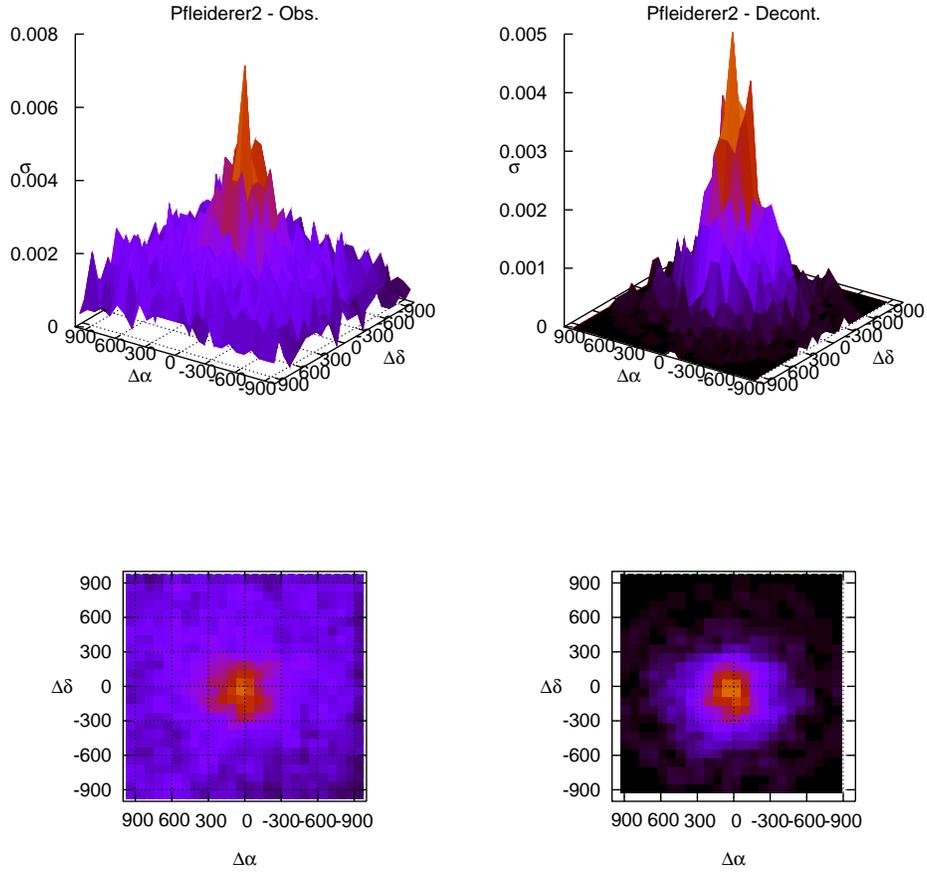}}}
\caption{Pfleiderer\,2: Surface density distribution built with the raw 
(left panels) and decontaminated photometry (right panels). $\sigma$ is 
given in units of $\rm stars~pixel^{-2}$; $\Delta\alpha$ and $\Delta\delta$
are in pixel. }
\label{fig3}
\end{figure}

In Fig.~\ref{fig3} we investigate the surface density ($\sigma$, in units of 
$\rm stars~pixel^{-2}$) distribution of Pfleiderer\,2 both with the observed (left panels) 
and decontaminated (right) photometry. We also include the respective isopleth surface maps
(bottom panels), which show a rather circular geometry in the stellar distribution around the
center of Pfleiderer\,2. As expected, the cluster is much more contrasted in the decontaminated 
data. 

\section{Color Magnitude Diagrams}
\label{CMD}

\begin{figure}
\centerline{
\includegraphics[angle=0,scale=.50]{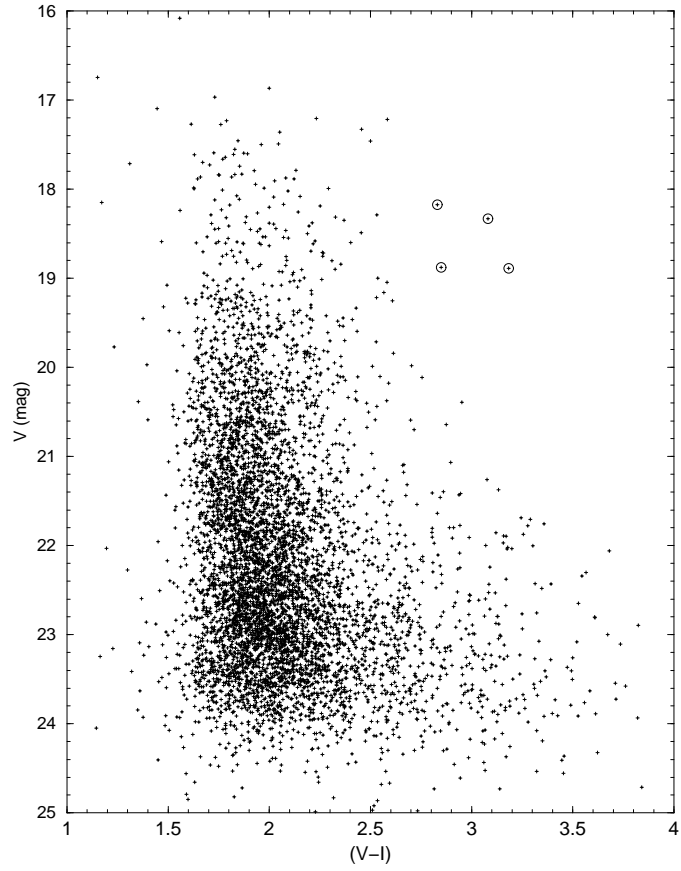}}
\caption{ V vs. V-I CMD of Pfleiderer\,2 for an extraction of r $<$ 1000 pixels 
or r $<$ 4.2' (essentially the full field). The bright and red stars (circled) 
suggest a metal-rich cluster.}
\label{fig4}
\end{figure}

Fig.~\ref{fig4} shows the $V$ vs. $V-I$ CMD of essentially the full field 
(r $<$ 1000 pixels, or r $<$ 4.2\arcmin) {\bf around Pfleiderer\,2, reaching 
the photometric limit of $V\approx24$. As anticipated in Sect.~\ref{Decont}, 
field contamination is dominant in this direction. A few bright and red stars 
show up at $V\approx18.5$ and $(V-I)\approx3$, which might suggest a metal-rich
cluster.}

\begin{figure}
\centerline{
\resizebox{10.0cm}{!}{\includegraphics[angle=-90,scale=.50]{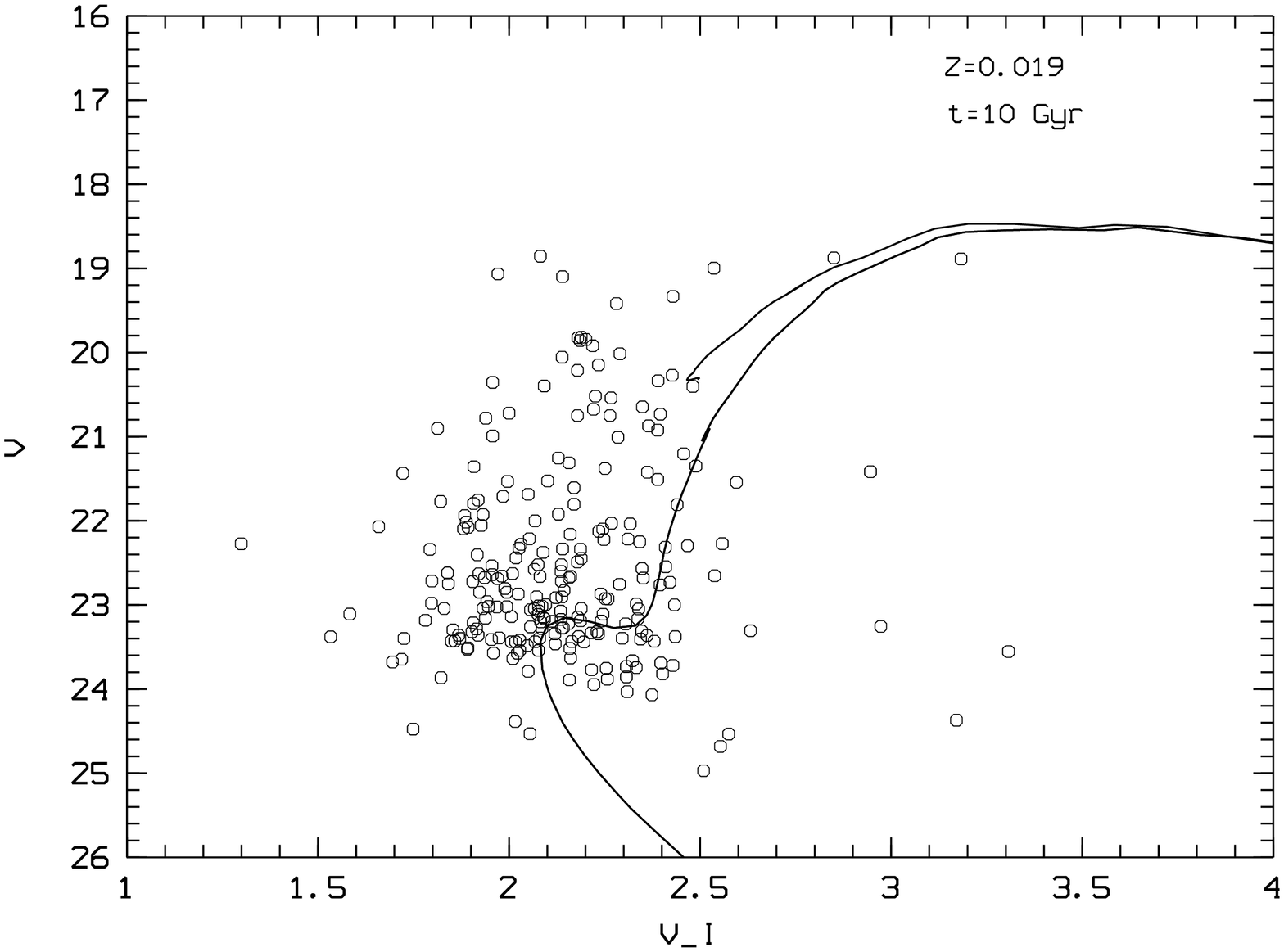}}
\resizebox{10.0cm}{!}{\includegraphics[angle=-90,scale=.50]{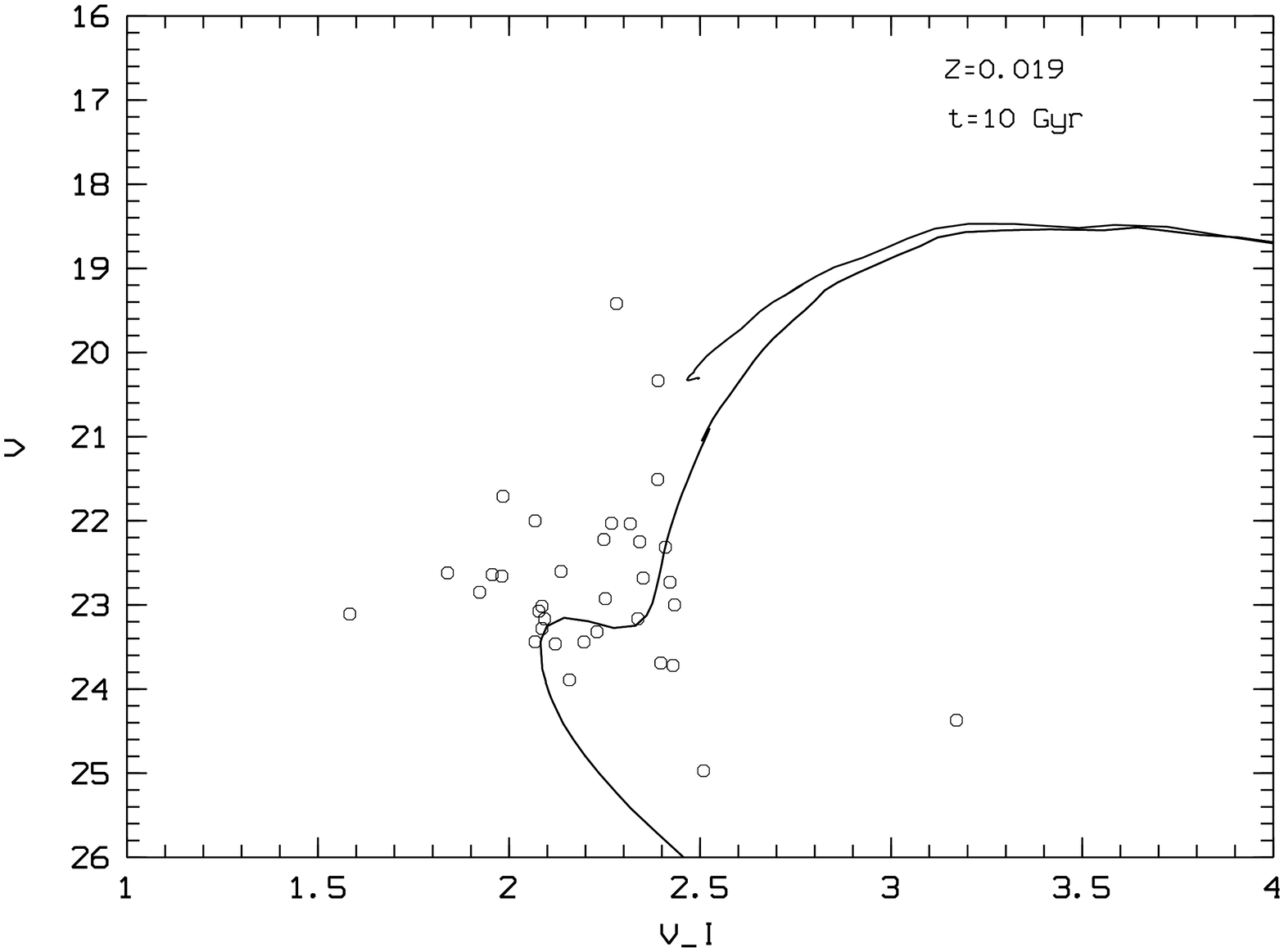}}
}
\caption{Solar metallicity Padova isochrone of 10 Gyr fit to the
V vs. V-I CMD of Pfleiderer\,2 for extractions of 30\arcsec\ (left) 
and 10\arcsec\ (right). }
\label{fig5}
\end{figure}

The fit of Padova isochrones (\citealt{GBBC00}) to the V vs. V-I CMD of Pfleiderer\,2
is shown in Figs.~\ref{fig5}a,b for extractions of r $<$ 10\arcsec\ and 30\arcsec.
A tilted Red Giant Branch (RGB) is seen, together with a clump of stars at $V=20.37\pm0.05$ 
and $(V-I)=2.4\pm0.1$, compatible with a red Horizontal Branch (HB). A tilted RGB is 
indicative of a high metallicity as discussed in \citet{OBB91}. We also note that the 
spread across the diagram is due to differential reddening.
 
The fit of the isochrones to the cluster sequences
in these Figs. required a high metallicity
 of Z=0.019 ([Fe/H]$\approx$0.0). It might be one of the most
metal-rich globular clusters in the Galaxy. 
The age estimated is 10$\pm$2 Gyr, compatible with a globular
cluster. 

This leads to a reddening of E(V-I)=1.34. Assuming
E(V-I)/E(B-V)=1.33 (\citealt{DWC78}), we get E(B-V)=1.01$\pm$0.15,
which corresponds to  A$_{\rm V}$ = 3.13. This reddening value
is compatible with the dust emission maps by \citet{Schl98},
of E(B-V)=1.15.
We derive an observed distance modulus of (m-M)$_{\rm V}$ = 19.2,
an absolute distance modulus of (m-M)$_{\rm \circ}$ = 16.07, and   
a distance from the Sun d$_{\odot}$ = 16.4$\pm$2 kpc.

  Assuming the conservative distance of the Sun to  the Galaxy center to be 
R$_{\odot}$ = 8.0 kpc (\citealt{Reid93}), the Galactocentric coordinates 
are X = 7.0 kpc (X $>$ 0 is on the other side of the Galaxy),
 Y = 6.1 kpc and Z = 2.7 kpc, and the distance to the Galactic
 center is R$_{\rm GC}$ = $9.7\pm2$\,kpc.

A metal-rich globular cluster located  far from the center
 is not a unique case.  Terzan\,7 with a 47 Tuc-like metallicity, in the
Sagittarius dwarf, is even farther from the Galactic center
 at R$_{\rm GC}$ = 15.9 kpc (\citealt{SW02}).
However, Pfleiderer\,2 is apparently not related to any
dwarf galaxy and is more metal-rich. 
The metal-rich cluster Palomar\,8 has [Fe/H]=-0.48 
and  R$_{\rm GC}$ = 5.6 kpc (H96), whereas Palomar\,11 has
[Fe/H]=-0.39 (H96) and R$_{\rm GC}$ = 9 kpc  (\citealt{Lewis06}).

Therefore, Pfleiderer\,2 has at least two other analog
metal-rich globular clusters located far from the bulge.
In case there is no relation to a dwarf galaxy, it might be
that these clusters are at apogalacticon, possibly tracing
such limit for the bulge.

\subsection{Decontaminated TNG photometry}

\begin{figure}
\centerline{
\resizebox{12.0cm}{!}{\includegraphics[angle=0,scale=.50]{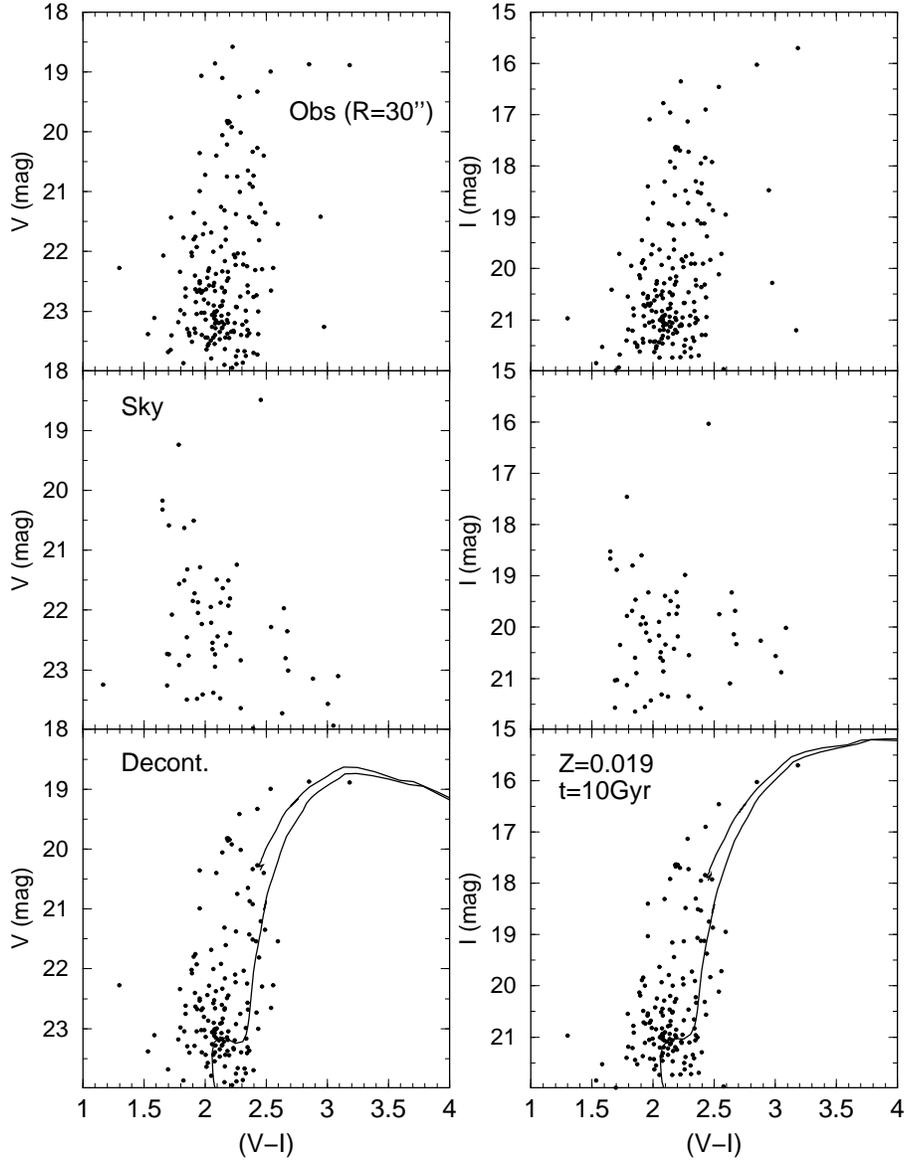}}}
\caption{ V vs. V-I and I vs. V-I CMD extraction of Pfleiderer\,2
for r $<$ 120 pixels (r $<$ 30\arcsec). Upper panels: observed; Middle panels:
equal area sky extractions; Lower panels: Decontaminated CMDs with
the isochrone solution from Fig.~\ref{fig5}. }
\label{fig6}
\end{figure}

\begin{figure}
\centerline{
\resizebox{12.0cm}{!}{\includegraphics[angle=0,scale=.50]{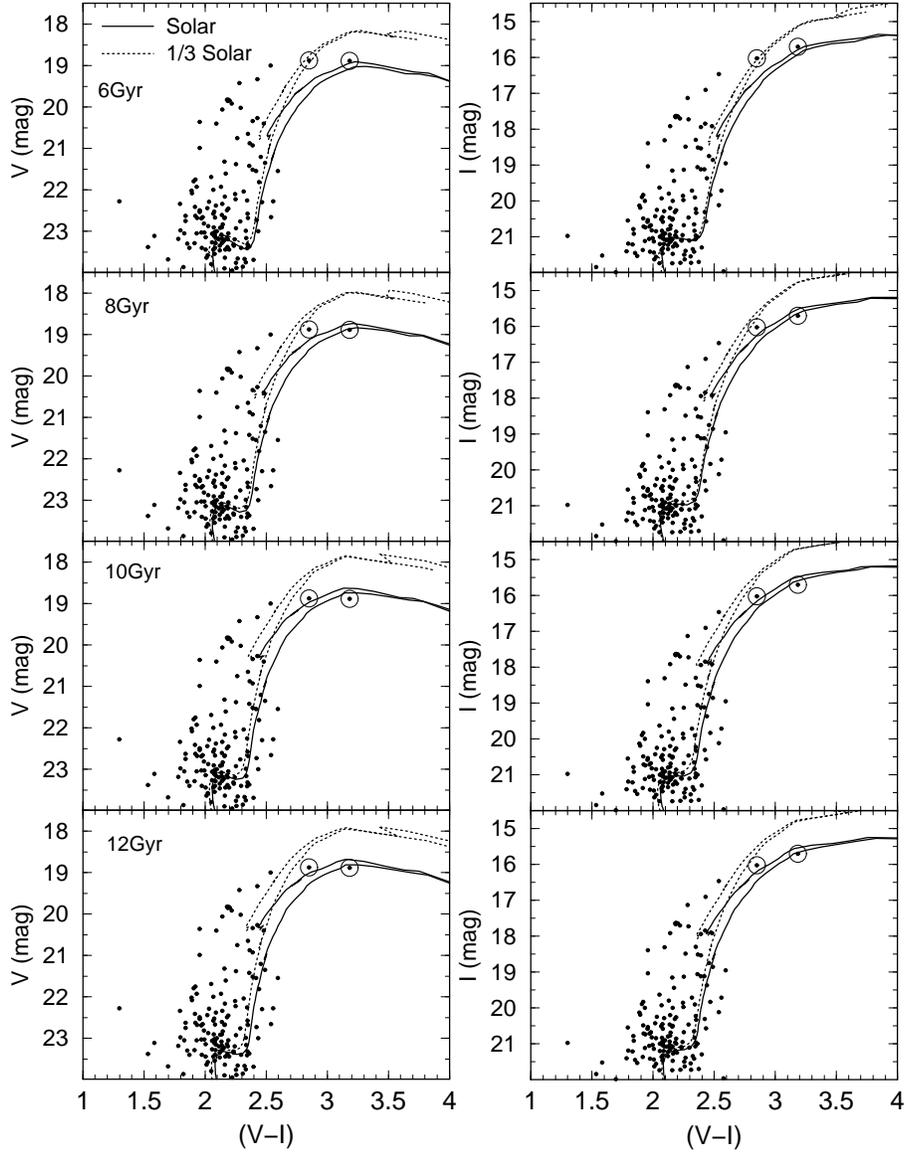}}}
\caption{Alternative age and metallicity solutions applied to the r $<$ 30\arcsec\
decontaminated CMD. Acceptable fits are obtained with nearly-solar metallicities 
and ages within 8-10\,Gyr. The bright and red stars that support the metal-rich
nature of this cluster are highlighted by circles.}
\label{fig7}
\end{figure}

We applied the decontamination tools to the TNG photometry, {\bf as described in 
Sect.~\ref{Decont}. } A mosaic of CMDs extracted within 120 pixels (r $<$ 30\arcsec) 
is given in Fig.~\ref{fig6}. The upper panels show the observed V vs. V-I and I vs. 
V-I CMDs, while the respective equal area sky extraction is given in the middle panels. 
The lower panels show the statistically decontaminated CMDs, derived according to 
Sect.~\ref{Decont} using as comparison field the wide ring located within 
$750<{\rm r(pixel)}<950$ ($3.13<{\rm r(\arcmin)}<3.96$). 

An RGB, the red HB and a populous turn-off are seen, features that are clearer in the 
decontaminated plot. These features are also seen in the V vs B-V subtracted diagram,
although not as deep as in V vs. V-I (Fig.~\ref{fig6}). {\bf At this point it is important
to recall that the two bright and red stars ($(V-I)\approx3$, $V\approx19$) have a
survival frequency of 100\%, with respect to the decontamination procedure 
(Sect.~\ref{Decont}). Besides, the redder one is located at about 14\arcsec\ from the
cluster center, while the other at about 29\arcsec. Taken together, these arguments
imply a high probability of both being cluster members, and suggest a metal-rich cluster.}

The CMD sequences seen in these decontaminated diagrams are similar to those obtained 
from central extractions (Fig.~\ref{fig5}a,b). {\bf As a caveat we note that the above
isochrone solutions place the turnoff at $V\approx23$ and $I\approx21$, which corresponds 
to about 1 mag brighter than the photometric limit, and probably close to the completeness
limit. However, the comparison with the field CMD (Fig.~\ref{fig6}) shows that the excess
of stars at $V\ga22.5$ is real.}

Finally, in Fig.~\ref{fig7} we test alternative isochrone solutions that explore a range 
of ages and metallicities. We test isochrones of 6, 8, 10 and 12\,Gyr, both of solar and 
1/3 solar metallicity. {\bf Assuming that the two bright and red stars are members, the 
metallicity turns out clearly constrained to nearly solar, since lower metallicities would 
require brighter GB stars. The 6\,Gyr isochrones, although producing a reasonable fit of the 
turnoff and RGB, fail to describe both bright and red stars simultaneously.} Although we 
adopted 10\,Gyr as the best age solution, a spread of $\pm2$\,Gyr is acceptable. In the case 
of 8\,Gyr, Pfleiderer\,2 might be one of those young globular clusters such as Palomar\,1 
(\citealt{Ros98}).

\section{Structure}
\label{Struc}

\begin{figure}
\centerline{
\resizebox{12.0cm}{!}{\includegraphics[angle=0,scale=.50]{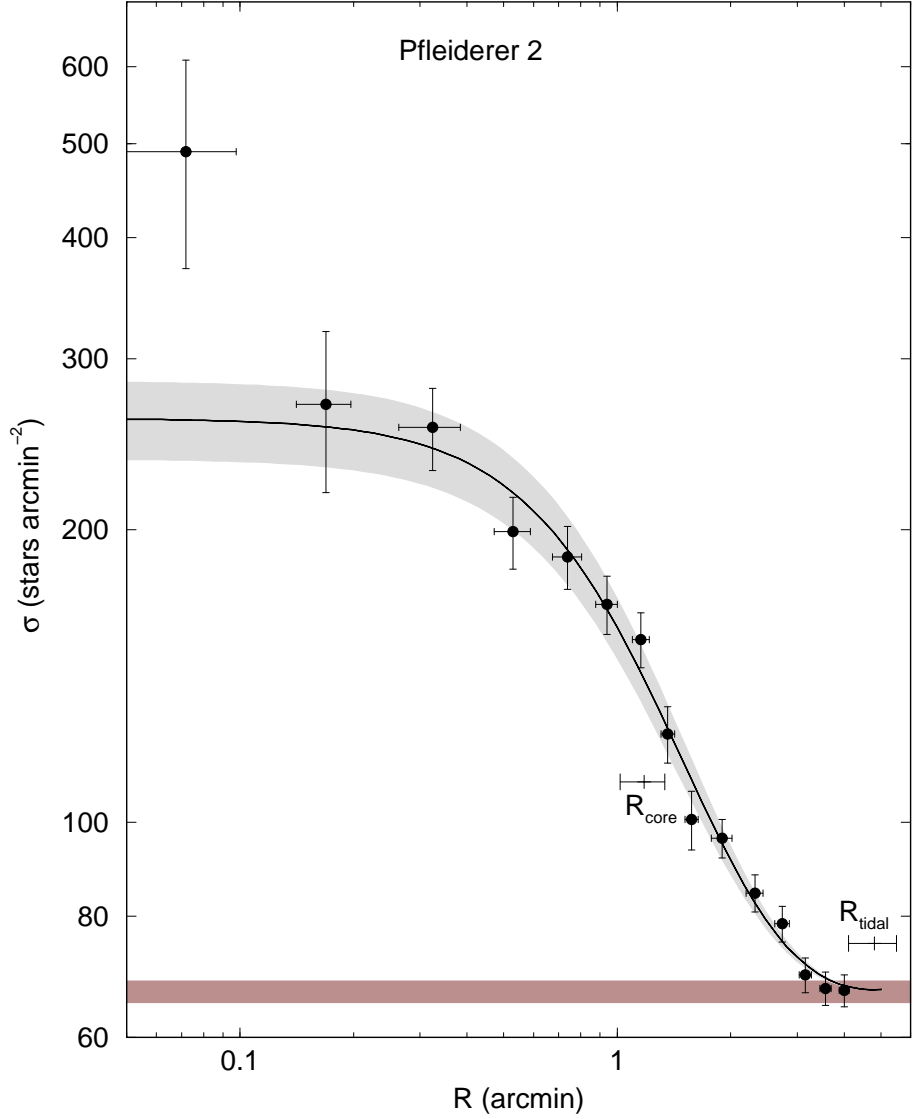}}}
\caption{Pfleiderer\,2 stellar radial density profile. The solid line corresponds 
to the best-fit 3-parameter King-like profile. The shaded area corresponds to the
1-$\sigma$ King fit uncertainty. Horizontal shaded bar is the background level. 
The cluster core and tidal radii are indicated.}
\label{fig8}
\end{figure}

In Fig.~\ref{fig8} we give the cluster stellar radial density distribution (RDP). The 
central parts of the cluster show evidence of a post-core collapse, as seen in a number
of other globular clusters, especially in the bulge (\citealt{TKD95}). Otherwise a smooth
fit with the 3-parameter King profile\footnote{Applied to star counts, this function is 
similar to that introduced by \cite{King1962} to describe the surface-brightness profiles 
of globular clusters.} is obtained throughout the rest of the cluster. This tentative fit 
provides a core radius of r$_{\rm c}(\arcmin) = 1.18\pm0.16$ ($\approx5.6$\,pc) and a tidal 
radius r$_{\rm t}(\arcmin) = 4.8\pm0.7$ ($\approx22$\,pc).

\citet{IHW83} determined tidal radii of globular clusters as a function of the estimated 
perigalacticon distances. Pfleiderer\,2 compared to their sample appears to be among the 
intrinsically smallest globulars, only comparable with the central bulge clusters NGC\,6522 
and NGC\,6528. In case the tidal radius is essentially determined by the perigalacticon tidal 
forces, the small size of this cluster suggests that it may reach  the outer bulge (see equation 
4 in \citealt{IHW83}).

On the other hand, the total absolute magnitude, obtained by adding the stellar 
fluxes for the stars brighter than V=24 present in the decontaminated CMD extracted 
within 700 pixels, amounts to M$_{\rm V}$= -2.5. This value would place Pfleiderer\,2 
among the least luminous globular clusters in the Galaxy, like Palomar\,clusters.
Therefore, if its mass is as low as that of Palomar\,clusters such as Palomar\,12 and 
Palomar\,13 (\citealt{GO88}; \citealt{ORS85}), estimated to be of $10^3$ to $10^4$ 
M$_{\odot}$, then a perigalacticon distance of 5 to 10 kpc is inferred. In this case,
it might be a low mass globular cluster that has been losing mass, or be an old open 
cluster, with a nearly circular orbit.

\begin{figure}
\centerline{
\resizebox{12.0cm}{!}{\includegraphics[angle=0,scale=.50]{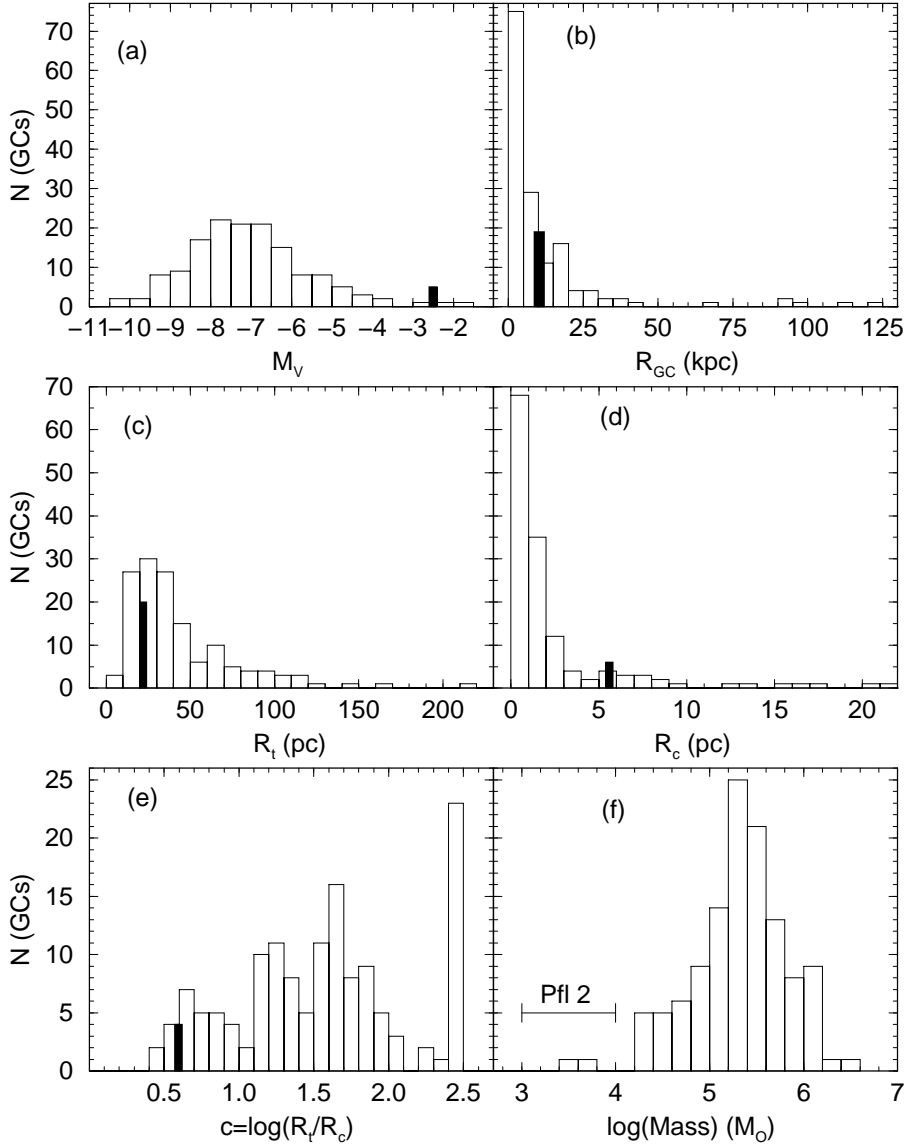}}}
\caption{Astrophysical parameters of Pfleiderer\,2 (black dash) are compared 
to those of the Galactic globular clusters (histograms). Panels show: the absolute
V magnitude (a), Galactocentric distance (b), King tidal (c) and core (d) radii,
concentration parameter (e), and mass (f). Histograms are built with H03 data, except 
the mass (f), taken from \citet{GO97}.}
\label{fig9}
\end{figure}

To put Pfleiderer\,2 into perspective, in Fig.~\ref{fig9} we compare some of the 
astrophysical parameters derived in this paper with those of the globular clusters in 
H96 (updated in 2003 - H03), together with the mass (panel d) taken from \citet{GO97}. 
As anticipated above, Pfleiderer\,2 is located in the low-luminosity tail of the globular 
cluster distribution (panel a) and not far from the Galactic center (b). The King tidal 
radius (c) is relatively small, consistent with its location in the Galaxy, while the 
King core (d), which encompasses the collapsed core (Fig .~\ref{fig8}), is clearly an 
upper limit. Disregarding the collapsed core, its King concentration parameter would 
suggest a relatively loose cluster (e). Undoubtedly, Pfleiderer\,2 is a low-mass 
cluster (f).

\section{Concluding remarks}
\label{Conclu}

The present observations deal with a faint star cluster that we identified as a globular 
cluster. Previously, Pfleiderer\,2 was classified as an open cluster. It is obscured in 
the optical, with $A_V=3.1$, and affected by differential reddening. Deep photometry
was collected with the 3.58m TNG telescope, which revealed the cluster nature, although 
deeper photometry would still be necessary for a final conclusion.

Isochrone fits are compatible with a nearly solar metallicity star cluster with age
within the range 8-12\,Gyr, which thus puts Pfleiderer\,2 among the few metal-rich 
globular clusters in the Galaxy. Larger apertures would be important to get deeper 
imaging, as well as high resolution spectroscopy to better derive its metallicity, 
abundance ratios, and orbital properties.

\begin{acknowledgements}
We thank an anonymous referee for suggestions.
We acknowledge partial financial support from the Brazilian agencies CNPq and 
Fapesp, and Ministero dell'Universit\`a e della Ricerca Scientifica e Tecnologica 
(MURST), Italy.
\end{acknowledgements}


%

\clearpage

\clearpage
\clearpage

\clearpage

\clearpage

\clearpage

\clearpage

\clearpage

\clearpage

\end{document}